\documentclass[prb,twocolumn,showpacs,10pt]{revtex4-1} 
\usepackage{amsmath,amssymb,graphicx}
\usepackage{todonotes}
\usepackage{subfigure}
\usepackage[export]{adjustbox}
\usepackage{array}
\usepackage[english]{babel}

\begin{document}

\title{Constructive feedback for the growth of laser-induced periodic surface structures}

\author{Jean-Luc D\'eziel}
\author{Joey Dumont}
\author{Denis Gagnon}
\author{Louis J. Dub\'e}
\email{Louis.Dube@phy.ulaval.ca}
\affiliation{D\'epartement de physique, de g\'enie physique et d'optique \\Facult\'e des Sciences et de G\'enie, Universit\'e Laval, Qu\'ebec G1V 0A6, Canada}
\author{Sandra Messaddeq}
\author{Youn\`es Messaddeq}
\affiliation{Centre d'Optique, Photonique et Laser, Universit\'e Laval, Qu\'ebec, Qu\'ebec G1V 0A6, Canada}

\begin{abstract}We study the formation and growth of laser-induced periodic surface structures (LIPSSs) with the finite-difference time-domain (FDTD) method. We use a recently proposed inter-pulse feedback method to account for the evolution of the surface morphology between each laser pulse sent to the surface of the processed material. This method has been used with an ablation-like mechanism, by removing material exposed to a light intensity higher than a given threshold. We propose an inverse mechanism, an expansion-like mechanism, able to grow structures that the ablation-like process cannot. This allows us to introduce the notions of constructive and destructive feedback and explains a strong contradiction between the standard Sipe-Drude theory and the experimental observations, i. e. the formation on metals of structures usually linked to wide band gap dielectrics.
\end{abstract}

\maketitle   

\section{Introduction}

The study of laser-induced periodic surface structures (LIPSSs) has started with the first experimental observation by Birnbaum \textit{et al}.\cite{Birnbaum1965} They are created by irradiating the surface of solid material (dielectrics, semi-conductors, metals)\cite{Driel1982} with an intense laser beam near the ablation threshold. LIPSSs are wavy nanometric structures characterized with both their orientation with respect to the incident wave polarization and periodicity $\Lambda$. A large variety of distinct LIPSSs morphologies\cite{Young1983,Wu2003,Wang2005,Vorobyev2008,Dufft2009,Rohloff2011,Bonse2012,Hohm2012} can be formed depending on the material and the laser properties.

We will consider two of these morphologies. The first type of structures to be considered are the most common and also the first ones to have been observed. They have an orientation orthogonal to the incident laser beam polarization and a periodicity close to its wavelength, $\Lambda \sim \lambda$. They are referred to as low spatial frequency LIPSSs (LSFLs) or as \emph{type-s} because of the sinusoidal dependency between the angle of incidence and their periodicity.

The second type of structures have an orientation parallel to the incident light polarization and a periodicity closer to $\Lambda \sim \lambda/\mathrm{Re}(\tilde{n})$, where $\tilde{n}$ is the complex refractive index. Since the real part of the refractive index can vary significantly during intense laser irradiation, the periodicity of parallel structures can cover a wide range. They are sometimes referred to as LSFLs or high spatial frequency LIPSSs (HSFLs) depending on their periodicity, or \emph{type-d} structures.

The currently accepted theory is the Sipe-Drude theory\cite{Sipe1983,Bonse2005,Bonse2009} which accounts for the interaction between an incident plane wave and surface scattered waves caused by the material's surface roughness. Sipe's analytical solutions of Maxwell's equations predict that a material with Re$(\tilde{n})>$ Im$(\tilde{n})$ (dielectrics) results in the formation of type-d structures and a material with Re$(\tilde{n})<$ Im$(\tilde{n})$ (metals or strongly ionized dielectrics) gives rise to type-s structures.

Agreement is quite good with the experiments on several points. On wide band-gap dielectrics at low laser fluence, type-d structures are usually observed.\cite{Hohm2012} On narrower band-gaps or at higher fluence, the resulting structures are rather of type-s because of the formation of dense plasma. On metals, type-s structures are observed most of the time, but in contrast to the Sipe-Drude theory, parallel structures have also been seen on metals and this observation is currently without explanation.\cite{Veld2010,Bonse2013}

Since the analytical method cannot predict the formation on metals of LIPSSs parallel to polarization, we turn to a numerical approach, the finite-difference time-domain (FDTD) method\cite{Yee1966,Skolski2012,Deziel2015} to investigate further. A feedback mechanism is also included to account for the growth of LIPSSs from one laser pulse to the next.\cite{Skolski2014} An extension to the model allows us to add an important notion, the distinction between constructive and destructive feedback in the LIPSSs formation process. This turns out to be crucial for the formation of type-d structures on metals. 

\section{Numerical model}

\subsection{FDTD and discretization}

\begin{figure}
\centering
\includegraphics[scale=0.2]{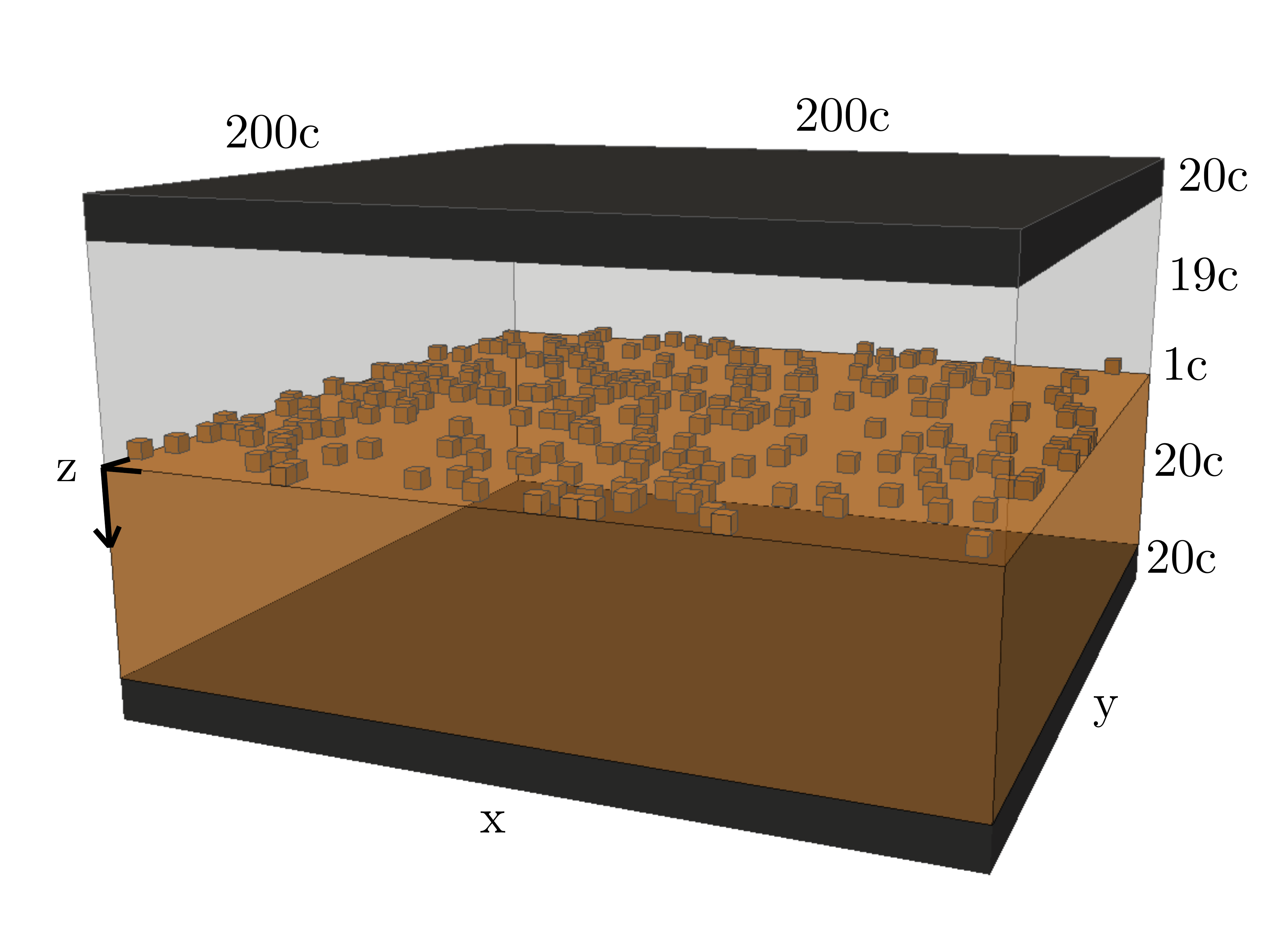}
\caption{Geometry used in the FDTD calculations. Material is represented in orange, void in light gray and PMLs in black. PMLs are also present at vertical boundaries (not shown).} \label{geometrySD}
\end{figure}

FDTD is a numerical method used to solve Maxwell's equations and is based on their spatial and temporal discretization. We first define a tridimensional domain, represented in Figure \ref{geometrySD}, in which half of the volume is occupied by the bulk of the material and the other half with void together with randomly distributed set of inhomogeneities positioned at the interface. Each inhomogeneity has the minimal volume allowed by the discretized space, i.e. one computational cell. Futhermore, 10\% of the cells in the layer above the bulk are randomly filled with material. Each spatial boundary is completed with perfectly matched layers (PMLs), 20 cells wide, to avoid any non-physical boundary reflections.

All spatial dimensions are normalized with respect to the incident light wavelength and the spatial discretization is $\delta_{x,y,z}=\lambda/20$ in a domain of 10$\lambda$ in $x$ and $y$ and 2$\lambda$ in $z$, excluding PMLs. The time dimension is normalized with respect to the time of a complete optical period $T$ and the temporal discretization is $\delta_t=T/40$, smaller than the maximal time step $\delta_{t,max}$ allowed by the stability condition of the FDTD method ($\delta_{t,max} = T/(20\sqrt{3})$ in our case). A plane wave is then propagated from the highest layer of void towards the $+z$ direction and through the material roughness. Simulations run for 7 full optical cycles and we extract the average intensity $I(x,y,z)=\langle |\vec{E}(x,y,z)|^{2} \rangle$ over the 3 final optical cycles.

\subsection{Drude model}

The optical properties of the material are calculated by the use of the Drude model to account for the dense plasma formation during the laser interaction.\cite{Bonse2009} The refractive index is defined as $\tilde{n}=\tilde{\epsilon}^{1/2}$ and the complex permittivity $\tilde{\epsilon}$ is
\begin{equation}
\tilde{\epsilon}=\epsilon + \tilde{\epsilon}_{\mathrm{Drude}}=\epsilon + (1- i\gamma/\omega) \left[- \frac{\omega_p^{2}/\omega^{2}}{(1+\gamma^{2}/\omega^{2})} \right],
\end{equation}
where $\omega$ is the laser frequency, $\gamma$ is the collision frequency of the free carriers defined as the inverse of the Drude damping time $\tau_D$ and $\omega_p$ is the plasma frequency defined as $\omega_p^{2}=e^{2}N_e/\epsilon_0 m^{*}_{\mathrm{opt}}m_e$. Finally, $e$ is the electric charge, $N_e$ is the free carrier density, $\epsilon_0$ is the free space permittivity, $m^{*}_{\mathrm{opt}}m_e$ is the effective optical mass of the electrons and $\epsilon$ is the material permittivity with no free carriers fixed at $\epsilon=4.84$ in this work.

\subsection{Feedback}

\begin{figure}
\centering
\includegraphics[scale=0.4]{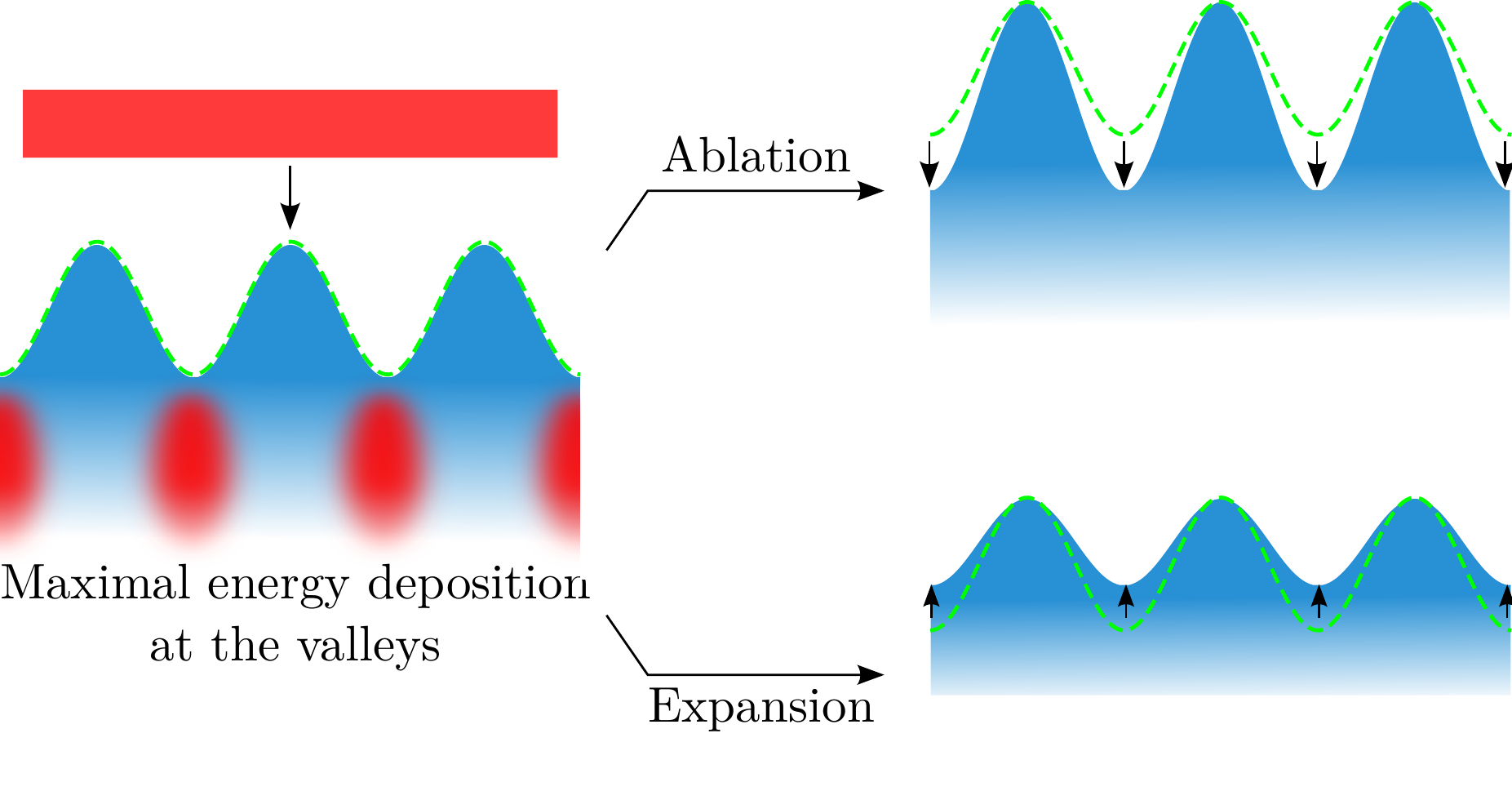}
\caption{Incident light is mostly directed to red spots. Amplitude growth (constructive feedback) is achieved by deeper ablation under the surface minima.} \label{feedbackMetal}
\end{figure}

\begin{figure}
\centering
\includegraphics[scale=0.4]{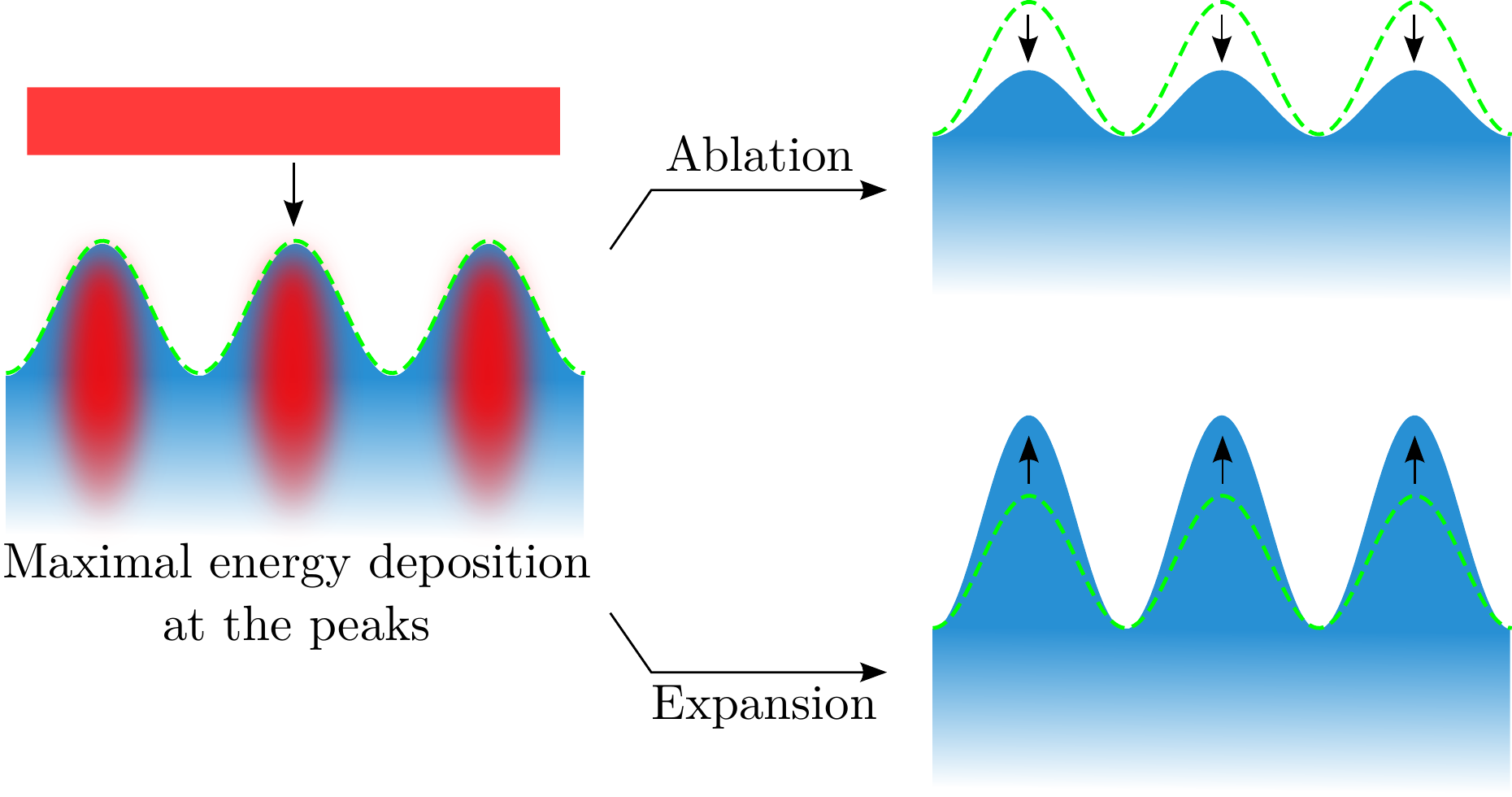}
\caption{Incident light is mostly directed to red spots. Amplitude growth (constructive feedback) is achieved by expansion of the surface maxima.} \label{feedbackDielectric}
\end{figure}

To account for the growth of LIPSSs between two laser pulses, an inter-pulse feedback mechanism recently proposed by Skolski \textit{et al.} is included.\cite{Skolski2014} After the first simulation, the time averaged light intensity $I(x,y,z)$ in the material is calculated and an isosurface $s(x,y,z)$ is traced along a threshold value of average intensity $I_{th}$. All the material above $s(x,y,z)=I_{th}$ has received more intensity than the threshold and is thus removed before starting a new simulation with the processed surface. This ablation-like process is repeated as many times as needed. The threshold is chosen to remove a precise average thickness of material $\Delta s$ after the first laser pulse and is kept constant for the subsequent pulses. $\Delta s$ is also in units of $\lambda$.

\begin{figure*}
\centering
\includegraphics[scale=0.98]{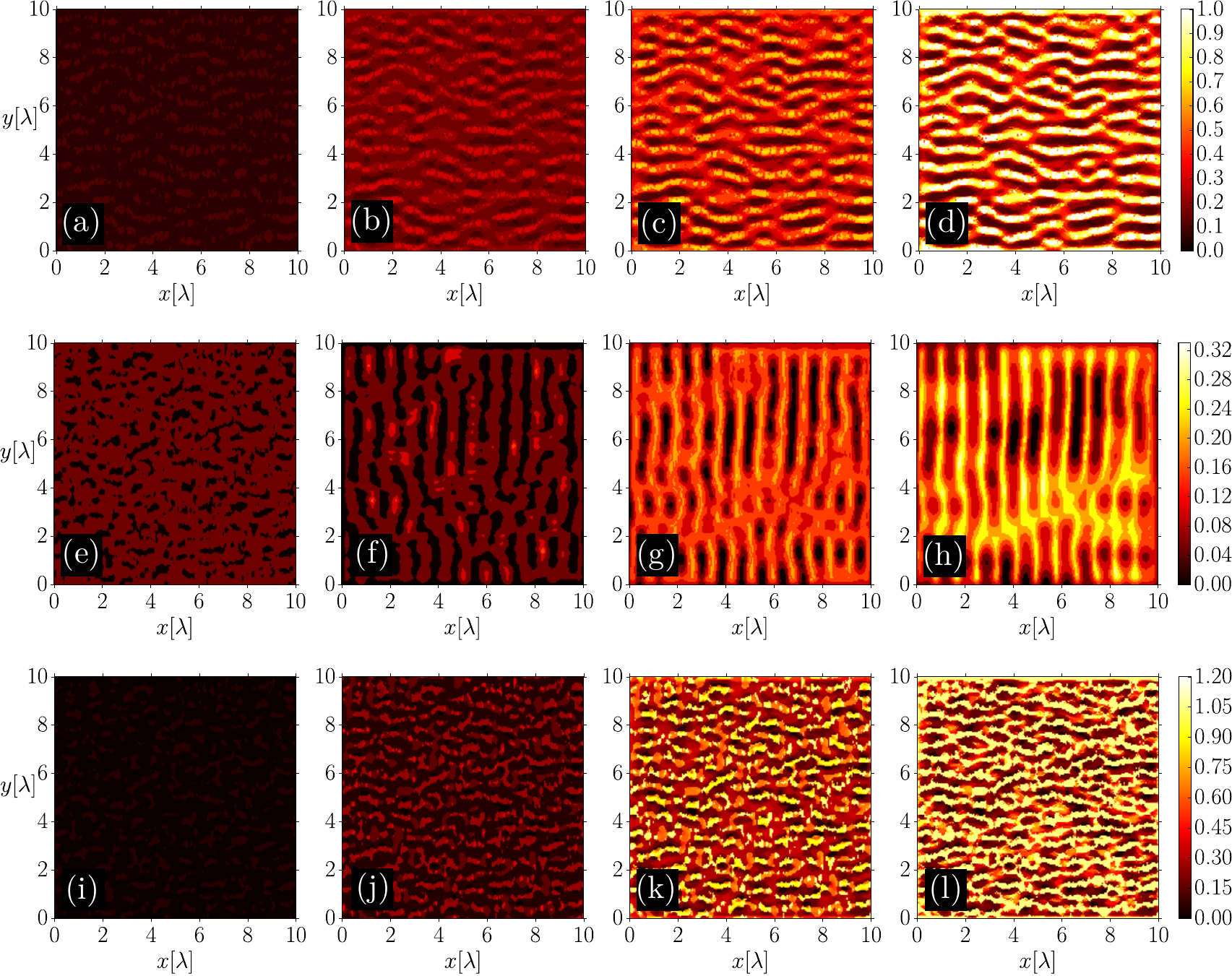}
\caption{Growth of periodic nanostructures with the inter-pulse feedback method. Subfigures (a) to (d) use parameters $(\omega_p/\omega,\gamma/\omega,\Delta s)=(1.7,1/16,-0.11)$ for pulse numbers 1,2,3 and 4 respectively. Subfigures (e) to (h) use parameters $(\omega_p/\omega,\gamma/\omega,\Delta s)=(6,1,0.16)$ for pulse numbers 1,3,5 and 9 respectively. Subfigures (i) to (l) use parameters $(\omega_p/\omega,\gamma/\omega,\Delta s)=(6,1,-0.16)$ for pulse numbers 1,3,5 and 9 respectively.} \label{growth}
\end{figure*}

Several distinct and complex mechanisms are responsible for the modification of the surface morphology in addition to ablation. For instance, photo-expansion will act differently on the surface because it will lead to expansion of the material where the light intensity is larger. We therefore introduce a second mechanism to account for any expansion-like mechanism with the same threshold method except that the surface modification is applied with the opposite sign. For example, if the difference between the actual surface position and the isosurface $s$ at a certain point is 5 cells, an ablation-like process will remove 5 cells of material and an expansion-like process will add 5 cells. We use negative values of $\Delta s$ to indicate an expansion-like mechanism.

Several scenarios can occur between two inter-pulse feedback cycles. When LIPSSs are already formed on a surface, the calculated mean intensity $I(x,y,z)$ can be maximal either at the surface minima (valleys) or at the surface maxima (peaks). If the energy is mostly deposited at the valleys, two scenarios illustrated in Figure \ref{feedbackMetal} are possible. If the chosen surface processing mechanism is ablation, the valleys get deeper and the total amplitude of the LIPSSs grows, leading to constructive feedback. In contrast, if the process is expansion-like, the valleys expand and the LIPSSs flatten as a result of destructive feedback.

In distinction, if the energy is mostly deposited at the peaks, as illustrated in Figure \ref{feedbackDielectric}, the opposite effects are expected. The ablation mechanism will erase the LIPSSs in destructive feedback and the expansion mechanism will lead to amplitude growth as a result of constructive feedback.

\section{Results \& discussion}

To test these assumptions, we first apply 9 full inter-pulse feedback cycles on a material with dielectric optical properties (Re$(\tilde{n})>$ Im$(\tilde{n})$) under the expansion-like mechanism. The parameters are $(\omega_p/\omega,\gamma/\omega,\Delta s)$ = $(1.7,1/16,-0.11)$, corresponding to $\tilde{n}=1.4+0.064i$ and results are shown in Figure \ref{growth}(a)-(d). The analytical solution of the Sipe-Drude theory predicts the formation of structures parallel to the polarization, in agreement with this simulation. Furthermore, our calculation confirms the illustration of Figure \ref{feedbackDielectric} whereby structures grow from one laser pulse to the next when expansion is applied, whereas they do not under ablation (not shown in Figure \ref{growth}).

The next two sets of calculations consider materials with metallic properties (Re$(\tilde{n})<$ Im$(\tilde{n})$). We again apply 9 full inter-pulse feedback cycles, but this time under the ablation-like mechanism with parameters $(\omega_p/\omega,\gamma/\omega,\Delta s)$ = $(6,1,0.16)$, corresponding to $\tilde{n}=2.14+4.21i$. The simulation results are shown in Figures \ref{growth}(e)-(h) and are in agreement with the Sipe-Drude theory both predicting type-s structures. The constructive feedback under ablation mechanism is again qualitatively displayed in Figure \ref{feedbackMetal}.

Now, with the same three parameters, $(\omega_p/\omega,\gamma/\omega,\Delta s)$ = $(6,1,-0.16)$, we apply an expansion-like mechanism to obtain Figures \ref{growth}(i)-(l). One clearly sees that structures with orientation parallel to the laser polarization are grown, a behavior not accounted for by the Sipe-Drude theory on metallic materials. This is in sharp contrast with the experimental fact that these types of structures are indeed observed on metals.\cite{Veld2010,Bonse2013} Our simulations suggest that the growth of type-d structures on materials with metallic optical properties is the result of energy deposition at the peaks with the ensuing constructive feedback for the developement of the maxima as illustrated in Figure \ref{feedbackDielectric}.

\section{Conclusion}

We have extended a recently proposed feedback-by-ablation method\cite{Skolski2014} to account for expansion-like mechanisms in the LIPSSs formation process. This allows us to simulate the growth of a larger class of nanostructure morphologies, including parallel ripples on dielectrics and metals.

We have thereby investigated the importance of constructive and destructive feedback. In particular, we have proposed an explanation for the growth of parallel structures (type-d) in the optically metallic regime, something not accounted for by the standard theory without feedback. 

The authors acknowledge the financial support form the Canada Excellence Research Chair in Photonics Innovations of Y. Messaddeq and the Natural Sciences and Engineering Research Council of Canada (NSERC). We also acknowledge computational resources from Calcul Qu\'ebec and the free software project Meep.

%
%

\end{document}